 \documentclass[final,times]{elsarticle}

\usepackage{amssymb}
\usepackage{color}

\usepackage{epstopdf}

\begin{document}

\begin{frontmatter}



\title{Bistable Auxetic Mechanical Metamaterials Inspired by Ancient Geometric Motifs}


\author[Harvard,McGill]{Ahmad Rafsanjani}
\ead{rafsanjani@seas.harvard.edu}
\author[McGill]{Damiano Pasini\corref{cor1}}
\ead{damiano.pasini@mcgill.ca}

\address[Harvard]{Harvard John A. Paulson School of Engineering and Applied Sciences, Harvard University, 29 Oxford Street, Cambridge, Massachusetts 02138, USA}

\address[McGill]{Mechanical Engineering Department, McGill University, 817 Sherbrooke Street West, Montr\'eal, Qu\'ebec H3A OC3, Canada}

\cortext[cor1]{corresponding author}

\begin{abstract}
\includegraphics [width=\textwidth]{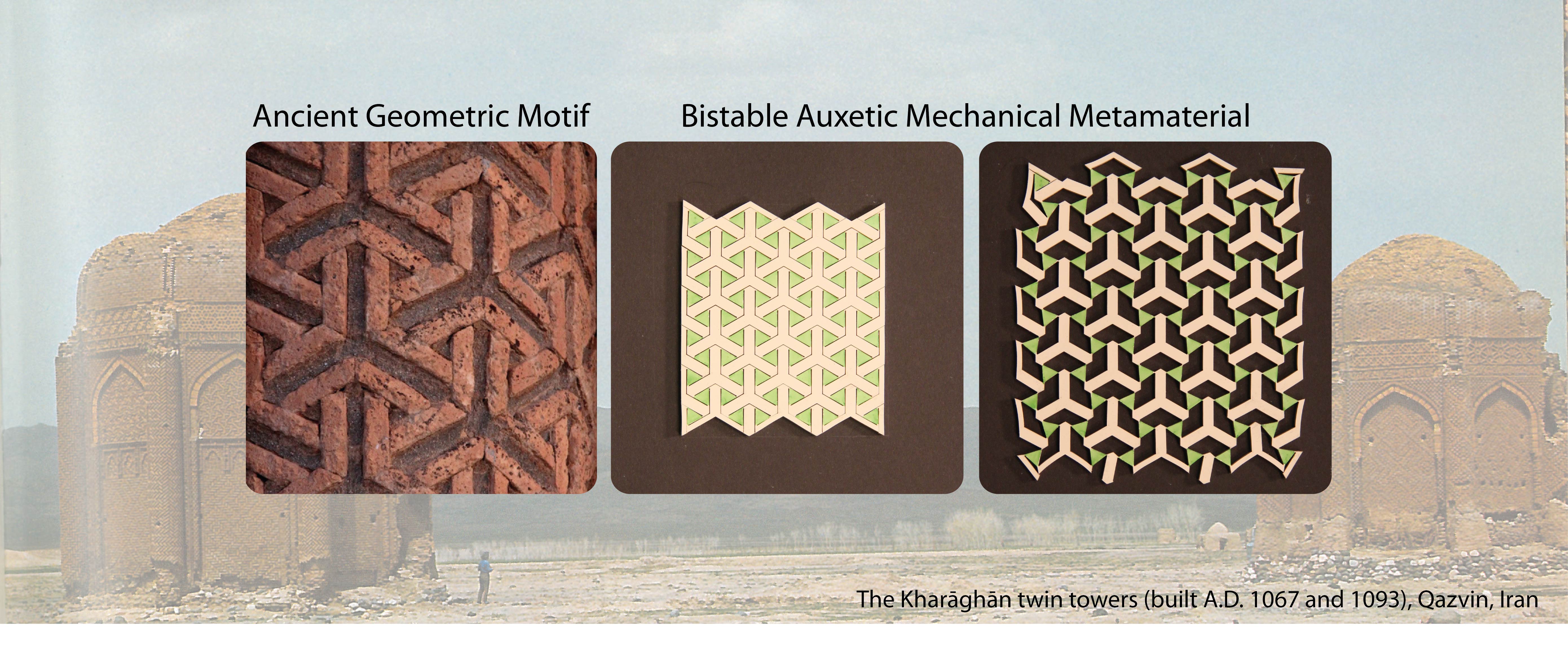}
Auxetic materials become thicker rather than thinner when stretched, exhibiting an unusual negative Poisson's ratio well suited for designing shape transforming metamaterials. 
Current auxetic designs, however, are often monostable and cannot maintain the transformed shape upon load removal. 
Here, inspired by ancient geometric motifs arranged in square and triangular grids, we introduce a class of switchable architectured materials exhibiting simultaneous auxeticity and structural bistability.  
The material concept is experimentally realized by perforating various cut motifs into a sheet of rubber, thus creating a network of rotating units connected with compliant hinges. 
The metamaterial performance is assessed through mechanical testing and accurately predicted by a coherent set of finite element simulations. 
A discussion on a rich set of mechanical phenomena follows to shed light on the main design principles governing bistable auxetics.

\end{abstract}


\end{frontmatter}


\section{Introduction}
\label{Introduction}

Mechanical metamaterials are designer matter with exotic mechanical properties mainly controlled by their unique architecture rather than their chemical make-up~\cite{Lee2012}. 
The Poisson's ratio, $\nu$, is the ratio between the transverse strain, $\varepsilon_t$, and the longitudinal strain, $\varepsilon_l$, in the loading direction ($\nu=-\varepsilon_t/\varepsilon_l$).
The mechanical behavior of materials with dissimilar Poisson's ratios can be very different~\cite{Greaves2011}.
The Poisson's ratio $\nu$ must satisfy $-1<\nu<0.5$ for isotropic 3D materials and $-1<\nu<1$ for isotropic 2D materials~\cite{Grima2005}.
This stability condition corresponds to the requirement that the stiffness tensor of the solid material must be positive definite~\cite{Lakes2008}.
For most solids, such as metals, polymers and ceramics, $\nu$ varies in a narrow band, i.e. $0.25<\nu<0.35$, which makes them contract transversely when stretched in the longitudinal direction.
In contrast, the reverse deformation is shown by materials with negative Poisson's ratio, known also as auxetics~\cite{Lakes1987, Keskar1992, Baughman1998} which for certain applications are desired~\cite{Ruzzene2003,Gatt2015,Taylor2014, Ali2014, Schaeffer2015}.
Here, our focus is on rotating unit auxetics; for a comprehensive overview on auxetic materials, readers are referred to several excellent review papers~\cite{Greaves2011, Evans2000, Yang2004, Alderson2007, Prawoto2012}.

Rotating unit auxetics are comprised of square and equilateral triangular units connected together by hinges at their vertices~\cite{Grima2000} (Fig.~S1).
If pulled in one direction, the collective motion of their rotating units induces the material to expand in the transverse direction, thereby giving rise to a Poisson's ratio of $-1$, a behavior similar to chiral auxetics~\cite{Prall1997, Scarpa2007} and some bar-and-joint lattices~\cite{Cabras2014}. 
One way to realize rotating unit auxetics in a planar sheet is to perforate it with elongated incisions along symmetric~\cite{Grima2010, Shan2015a}, randomly oriented~\cite{Grima2015}, or fractal patterns~\cite{Gatt2015, Cho2014}; in these cases, the resulting auxetic behavior can be isotropically tuned at targeted expandability, a feature that is an asset for highly flexible and stretchable devices. 
Very recently, nanoscale rotating unit auxetics are realized by self-assembly of 2D protein lattices~\cite{Suzuki2016}.
Also, anisotropic rotating unit auxetics can be created from connecting squares and rectangles with dissimilar size~\cite{Grima2011}.
Moreover, {\it origami}~\cite{Silverberg2014, Silverberg2015, Dudte2016} and {\it kirigami}~\cite{Shyu2015, Blees2015, Eidini2015}, respectively known as the arts of folding and cutting of paper, provide an alternative tool for designing auxetic metamaterials.
The rigid origami analogs of rotating unit auxetics are realized in Ron Resch's fold patterns~\cite{Tachi2013}.
Other examples are the well-known Miura-ori origami fold~\cite{Schenk2013, Silverberg2014} and its kirigami derivatives~\cite{Eidini2015, Eidini2016}, all exhibiting in-plane negative Poisson's ratio.

Shape transformations in auxetic materials can be programmed by exploiting the nontrivial deformation modes pertinent to elastic instabilities~\cite{Bertoldi2010, Overvelde2012, Babaee2013, Wu2015}.
However, achieving shape alterations that are robust and stable is challenging, since a pre-stressed state has to be maintained in the structure~\cite{Florijn2014}.
A promising alternative to by-pass this requirement is offered by materials and structures that can travel between stable states if needed~\cite{Qiu2004, Guest2006, Pandey2014}. 
Although numerous monostable auxetic designs exist, bistable auxetics are not fully explored yet, with solely very few origami designs~\cite{Silverberg2015,Yasuda2015}; such development is thus the focus of the current investigation, where the novel contribution lies in the integration of these two properties in one monolitic planar material, which is desirable in several applications such as medical stents and resizable screens.

\begin{figure} [t]
\centering
\linespread{1.0}
\includegraphics [width=\textwidth]{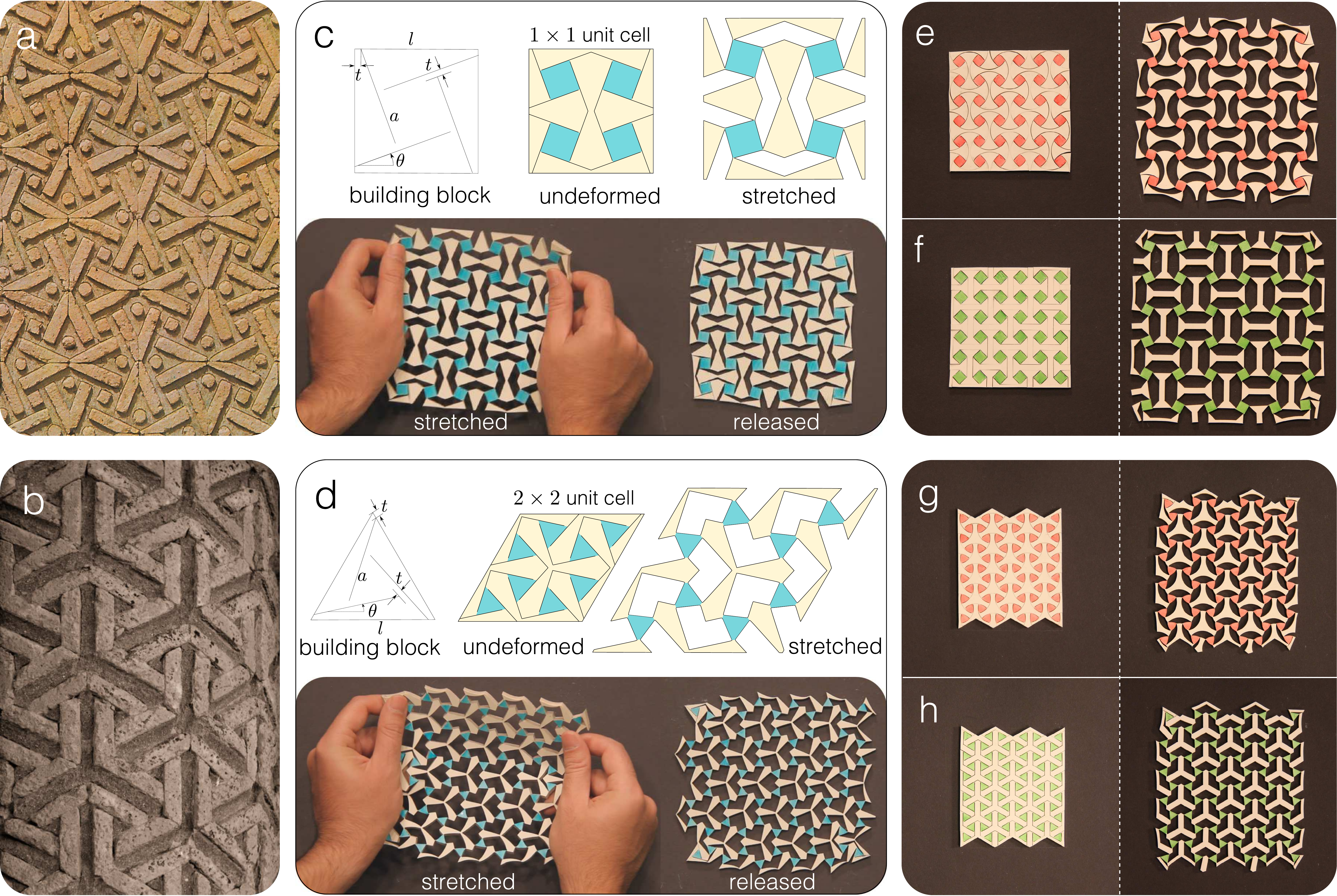}
 \caption{Bistable auxetics.  
{\bf a, b}, 
Brick decorations based on square and triangular motifs in the {\em Kharr{\={a}}q{\={a}}n} twin tomb towers in western Iran.
{\bf c, d}, The building block and the corresponding unit cell in its undeformed and deformed state comprises square and triangular rotating units with tilted cut motifs, which can maintain the stable deformation pattern after the load release. 
{\bf e-h}, Undeformed and deformed states of bistable auxetics with square and triangular rotating units for circular and parallel cut motifs.
See Movie S1 for further demonstration.}
\label{Fig1}
\end{figure}

\section{Bistable Auxetic Mechanical Metamaterials}

Our design is founded on perforated rotating unit auxetics.
Along this path, we take inspiration from geometric motifs in ancient architectures and present a concept for the design of planar bistable auxetics that is generic and readily applicable to other patterns while can be manufactured through simple cuts at various length-scale. 
We should note that this is not the first time that geometric art has had an impact on materials science; in fact, it was used, for example, to study quasicrystal patterns~\cite{Lu2007}.
We recognize that other geometric patterns predate Islamic era and can provide equivalent source of inspiration. 
In this work, we inspired by two brick architectures found in the {\em Kharr{\={a}}q{\={a}}n} twin towers constructed during the {\em Seljuk} era (circa 1067 and 1093) in Iran  (Fig.~\ref{Fig1}a and~\ref{Fig1}b)~\cite{Seherr1968}.
In particular, we focus on patterns comprising four (or three) mutually intersecting lines nesting a square (or equilateral triangle) unit within a square (or triangular) module (Fig~\ref{Fig1}c and~\ref{Fig1}d). 
Several variants can be created by merely altering an angle or curving a straight line in the repeating unit of these patterns.
We fabricated square and triangular specimens (Fig.~\ref{Fig1}c-h) by perforating three different cut motifs, namely tilted (blue), circular (red) and parallel (green) cuts into sheets of natural latex rubber (Fig. S2) using a laser cutter.
The length of the rotating unit was kept half of the length of its building block ($a/l=1/2$) where in all samples $l=20$~mm. 
The cuts are slightly shorter than the edge lengths, thereby leaving a finite-size ligament between units with size of $t=1$ mm.
We verified that the hinges are sufficiently thick to be manufactured by laser cutting without introducing major defects in the ligaments while being resistive to fracture under repeated loading.
The above condition implies that $\theta\simeq12^\circ$ for tilted triangular and $\theta\simeq20^\circ$ for tilted square motifs.
(Fig~\ref{Fig1}c, ~\ref{Fig1}d and Supplementary Materials).

\section{Results and Discussions}

We examined the performance of bistable auxetics by pulling apart the opposite edges until they were entirely expanded. 
The specimens underwent several metastable states before being fully developed, and retained their deformed shape even when the load was released (Fig.~\ref{Fig1}c-h and Movie S1).
Being monolithic with the rest of the structure, the flexure hinges bent and induced a relative rotation between neighboring units; as a result, snap-through instabilities occurred, a phenomenon that brought together auxeticity and structural bistability.
In contrast, conventional rotating unit auxetics resumed to their undeformed configuration upon load removal (Fig.~S1 and Movie S1).

\begin{figure} [t]
\centering
\linespread{1.0}
\includegraphics [width=\textwidth]{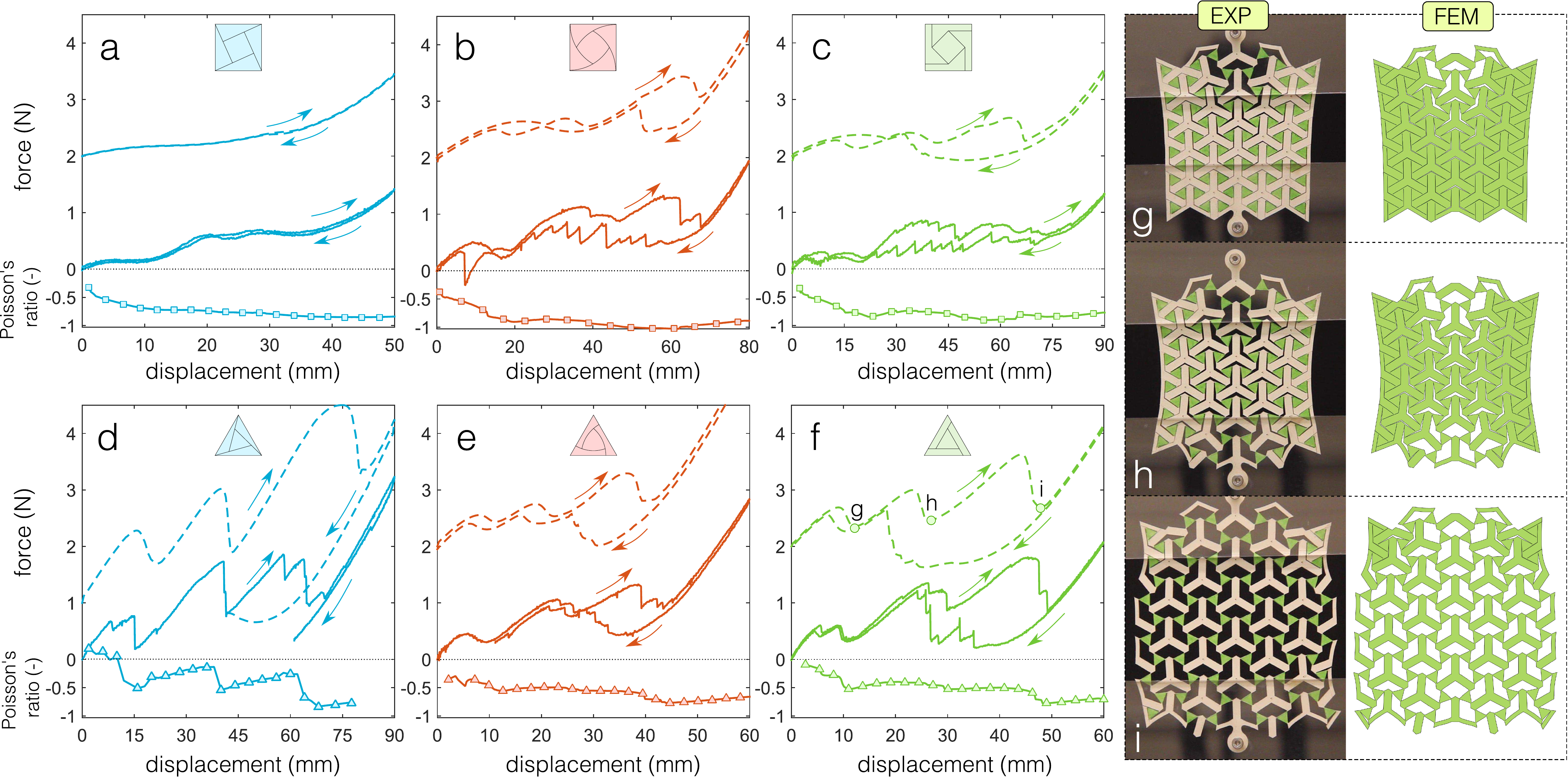}
 \caption{Mechanical response of bistable auxetics to uniaxial loading.
{\bf a-c}, Experimental (solid line) and FE (dashed lines, shifted vertically for readability) force-displacement curves during loading (tension) and unloading (compression) and experimentally characterized Poisson's ratio (symbols) of square bistable auxetics ($a/l=1/2$, $t/l=0.05$) respectively for tilted (blue), circular (red) and parallel (green) square cut motifs.  
{\bf d-e}, Similar results as above for triangular bistable auxetics.
Note that the numerical curves are shifted vertically for better readability.
{\bf g-i}, Comparison of experimental and FE deformed shape of a triangular design specimen with parallel cut motifs at selected displacements shown in diagram f.}
\label{Fig2}
\end{figure}

To further study the mechanical behavior of bistable auxetics, we measured their response under uniaxial tensile loading. 
The experiments revealed that an initial linear region caused by small bending of the ligaments is followed by a serrated force-displacement curve that accommodates most of the deformation.
The saw-tooth serrations and the resulting negative stiffness behavior are caused by snap-through buckling, from which the metamaterial bistability originates~\cite{Moore2006, Restrepo2015}.
At full extension, the response is governed by stretch dominated deformation of the compliant hinges (Fig.~\ref{Fig2}a-f and Movie S2).
The force-displacement curves taken during loading and unloading cycles exhibit a large hysteresis loop (except for the tilted square pattern), although the response is fully reversible and the material can fully recover its undeformed configuration upon unloading.
Moreover, the complex behavior of bistable auxetics is reasonably reproducible under cyclic loading, with response that shows a merely weak rate dependency (Fig.~S4).
The Poisson's ratio is negative through the full range of extension and approaches  -1 upon the specimen full expansion (Fig.~\ref{Fig2}a-f).

\begin{figure} [t]
\centering
\linespread{1.0}
\includegraphics [width=\columnwidth]{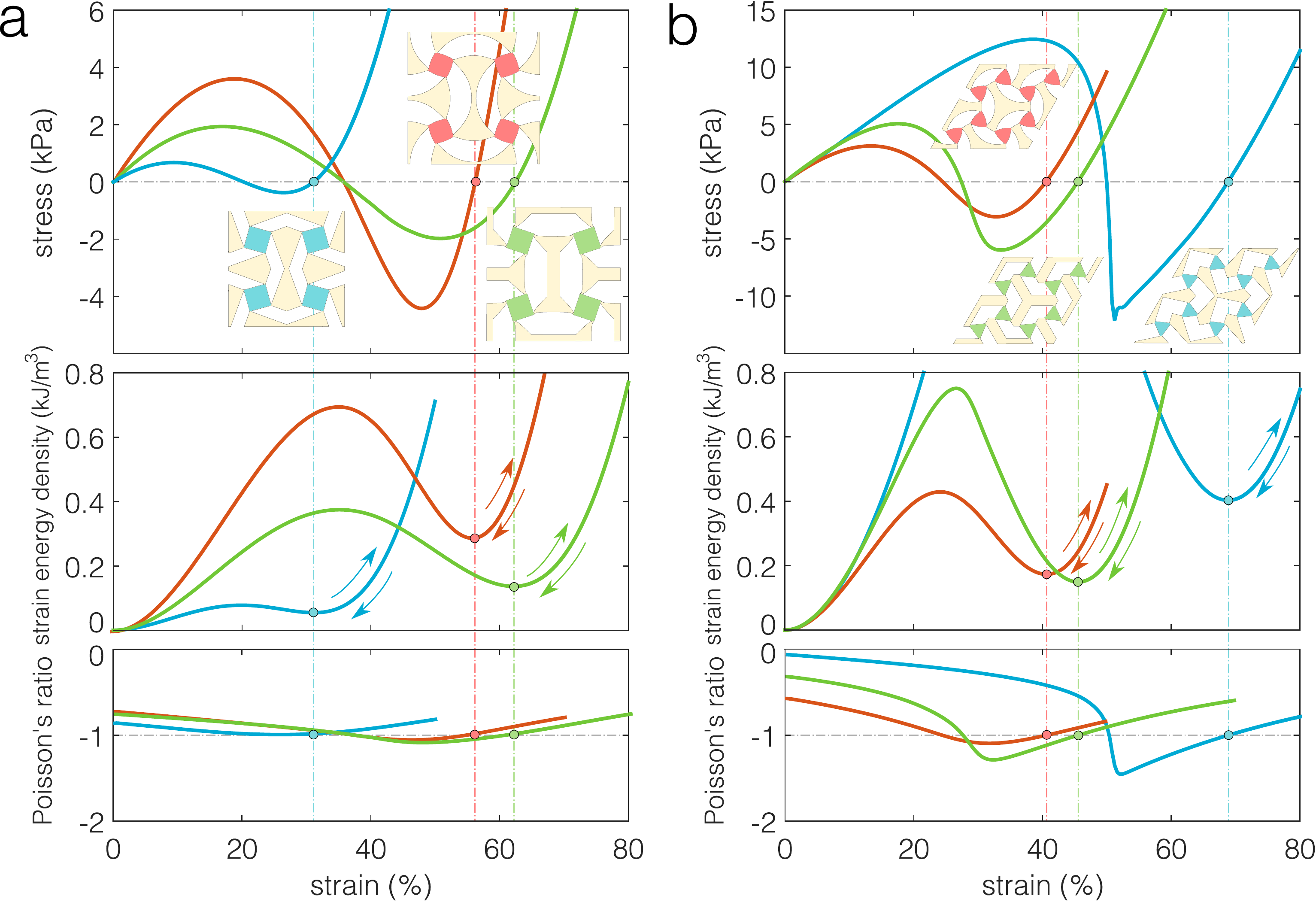}
 \caption{{\bf Mechanical response of unit cells obtained by FE simulations}.
{\bf a, b}, Stress, strain energy density and Poisson's versus strain respectively for square and triangular unit cells ($a/l=1/2$, $t/l=0.025$) with tilted (blue), circular (red) and parallel (green) cut motifs calculated by FE simulations under periodic boundary conditions.
The unit cells are subjected to uniaxial extension until they are fully expanded.
Upon load removal, the unit cells return to stable states (whose unit cells are shown in the insets) of local minimum energy (markers) where the Poisson's ratio equals to -1.}
\label{Fig3}
\end{figure}

A nonlinear FE analysis was performed to further explore the rich mechanical phenomena observed in the experiments (Supplementary Materials).
The force-displacement curves obtained by finite size FE models are quantitatively in good agreement with the measurements reproducing the key trends observed in the experiments (Fig.~\ref{Fig2}a-f, where dashed curves are offset along the y-axis for readability; see Movie S3).
By introducing contact, we assure that there is no overlap between the internal elements as observed in experiments.
Despite the simplified contact model used in our FE simulations, the predicted deformed shapes compare very well with those observed in the experiments (Fig.~\ref{Fig2}g-i).
The deformation remains in-plane for all investigated patterns except for the triangular design with tilted cuts.
In this case, the experiments showed that the metamaterial favors out-of-plane deformation before reaching full deployment, a response that is found to be energetically less expensive than the in-plane deformation enforced in the 2D model.
Consequently, the rotating units entangle if the out-of-plane deformation is impeded.
To gain further insight into the energetics of bistable auxetics while avoiding finite size effects, we simulated the response of a periodic unit cell subjected to uniaxial tension and released after full expansion.
We simultaneously tracked the engineering stress-strain relationship, the strain energy density and the Poisson's ratio of the material (Fig.~\ref{Fig3}).
Following a linear region, the stress-strain curve becomes strongly nonlinear and the load drops below zero before returning to positive values. 
It appears that the energy landscape of the system possesses a local minimum at a non-zero deformation which is pertinent to the second stable state of the material.
The Poisson's ratio is negative for all the material designs through the full range of the applied stretches, and interestingly it equals to -1 at the second stable state. 

\begin{figure} [t]
\centering
\linespread{1.0}
\includegraphics [width=0.9\textwidth]{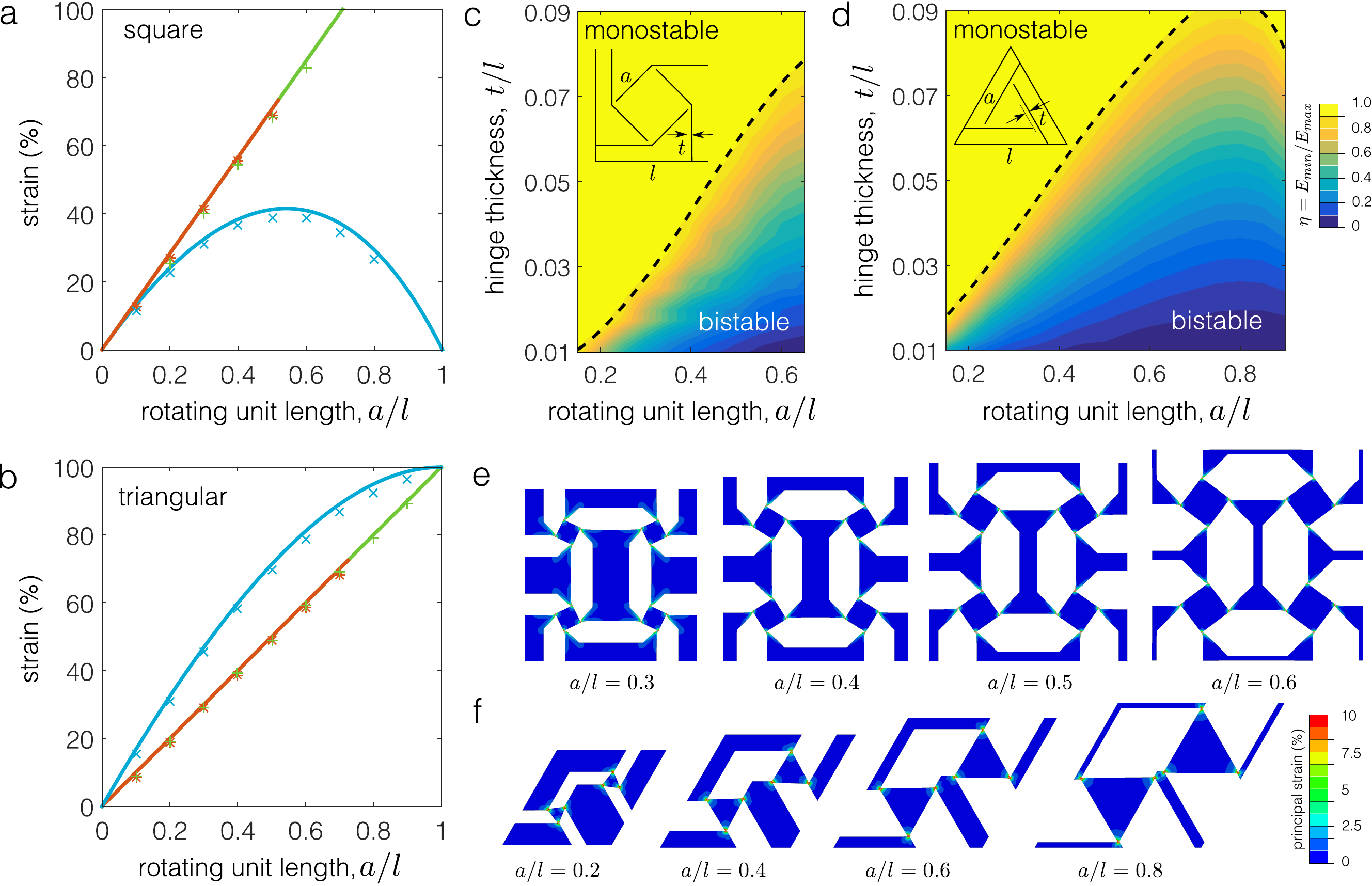}
 \caption{ Bistability.
{\bf a,b}, Geometric predictions (solid lines) of deformed stable states of metamaterials, respectively for square and triangular patterns with tilted (blue), circular (red) and parallel (green) cut motifs, compare well with FE simulations (symbols) for very thin hinges (here $t/l=0.005$).  
{\bf c, d}, Phase diagrams obtained by FE simulations respectively for square and triangular metamaterial designs with parallel cut motif show how metamaterial bistability is controlled by the normalized hinge thickness $t/l$ and the length of the rotating units $a/l$.
The contours show the ratio of the local minimum (stable deformation) to the peak (barrier) of the strain energy $\eta=E_{min}/E_{max}$. 
For $\eta<1$, the metamaterial is structurally bistable. 
{\bf e, f}, Stable deformed states of square and triangular bistable auxetics with parallel cut motif at selected lengths of the rotating units for $t/l=0.02$.
Contours of principal strains show that the deformation is localized at the hinges while both rotating and non-rotating units are undeformed.}
\label{Fig4}
\end{figure}

A simple kinematic analysis suggests that the two isolated states corresponding to the undeformed and fully deployed configurations of the metamaterial are permissible under the point hinge assumption (Fig.~\ref{Fig4}a and~\ref{Fig4}b and Supplementary Materials).
In practice, the flexures have finite thickness resisting the deformation induced by the rotational stiffness of the ligaments, and hence, not all the identified configurations are essentially bistable.
The compliance and expandability of the designed metamaterials can be tuned by controlling the basic cut profile of their building blocks.
This prospect opens up multiple avenues for shape optimization of cut profiles for a desired performance~\cite{Clausen2015}. 
We also examined how the structural bistability of the metamaterials relies on the hinge thickness $t/l$ and the width of the rotating unit $a/l$; 
we did so by calculating the ratio of the local minimum strain energy $E_{min}$ pertinent to the deformed stable state of the metamaterial to its strain energy barrier $E_{max}$, i.e. $\eta=E_{min}/E_{max}$ (Fig.~S7a).
This spectrum shed light on the appearance of transition states in the parameter space, which shows bistability for $\eta<1$ and metastable behavior for $\eta=1$ (Fig.~\ref{Fig4}c,~\ref{Fig4}d and Fig.~S7).
The distribution of the principal strains at the stable deformed state shows that the strain localizes mainly at the hinges with negligible deformation in the rest of the structure (Fig.~\ref{Fig4}e and~\ref{Fig4}f).

It is notable that the performance of bistable auxetic metamaterials are comparable to that of {\em origami twist tessellations} which are repeating patterns of pleats and twists folded from a single sheet of paper~\cite{Gjerde2008}.
In particular, certain square and triangle origami twists are geometrically akin to our metamaterial designs (Ref.~\cite{Gjerde2008} and Fig.~S8).
The switchable expansion of the structures described in this study is achieved by the flexural deformation of the ligaments in sheets perforated with simple cut motifs.
However, folding and unfolding of origami tessellations are sometimes laborious and lengthy to realize, since the crease patterns of origami twists are mathematically unfoldable and the bistable unfolding of each twist requires the finite bending of its facets~\cite{Silverberg2015}.
Recently, an origami-based metamaterial is proposed by Yasuda and Yang based on 3D assembly of the Tachi-Miura polyhedron (TMP)~\cite{Yasuda2015}. 
For certain parameters, the TMP metamaterial exhibits negative Poisson's ratio and structural bistability, however it bears key differences with respect to our design. 
Our proposed design is 2D suitable for planar applications whereas the TMP structure is 3D.
The Poisson's ratio of the TMP structure is anisotropic and varies from negative to positive values depending on the applied strain. In contrast, the Poisson's ratio of our proposed bistable auxetic remains negative and for certain cut patterns is approximately -1. 
The fabrication of a TMP structure requires to assemble individual origami elements, whereas our 2D bistable auxetic is monolithic and can be created from a single sheet by prescribing cuts with different motifs.

\section{Concluding Remarks}

In conclusion, we presented a novel class of planar mechanical metamaterial exhibiting a switchable expandability.
The combined structural bistability and negative Poisson's ratio in two dimensions has not been previously observed in perforated auxetics~\cite{Gatt2015, Shan2015a, Grima2015, Cho2014} and bistable metamaterials~\cite{Duoss2014, Shan2015b, Rafsanjani2015, Restrepo2015}, thus our work adds another dimension to the design space of architected soft materials. 
Expandability, stiffness and bistability of metamaterials can be conveniently controlled by varying the geometry of the basic cut profiles. 
This strategy can be exploited to design new deployable structures that do not require load control after deployment, flexible devices, medical implants to treat stenotic lesion in body vessels~\cite{Gatt2015, Ali2014}, as well as wearable skin sensors with tunable mechanical properties.

\section*{Acknowledgements}
A.R. acknowledges the financial support provided by {\em Swiss National Science Foundation} (SNSF) under grant no. 164648.
We thank Hang Xu for technical support and Shima Asa for assisting in filming and photography.


\appendix
 
 \section{Supplementary Materials}
Supplementary materials related to this article can be found online.




\end{document}